\documentclass[aps,pre,10pt,notitlepage,nofootinbib,showkeys,showpacs,twocolumn]{revtex4}
\usepackage{amstext,amsmath,amssymb,amsfonts,bbm}
\usepackage[latin1]{inputenc}
\usepackage{fancyhdr}
\usepackage{graphicx}
\usepackage{hyperref}
\usepackage{epstopdf}
\usepackage{amsthm}
\usepackage{amssymb,amsmath}
\usepackage{pstricks}
\usepackage{tocvsec2}
\usepackage{subfigure}

\usepackage{color}

\def\beq{\begin{equation}}
\def\be{\begin{equation}}
\def\ee{\end{equation}}
\def\bes{\begin{eqnarray}}
\def\ees{\end{eqnarray}}

\renewcommand{\u}{\mathfrak{u}}

\def\f{\frac}

\def\pp{\partial}

\def\calA(\Gamma,G){{\mathcal A}}

\def\u{\underset}

\theoremstyle{definition}
\theoremstyle{definition}
\theoremstyle{definition}
\theoremstyle{definition}
\theoremstyle{definition}
\theoremstyle{definition}

\begin{document}
\maxtocdepth{subsection}

\title{\large \bf Tailoring diffusion in analogue spacetimes}

\author{Matteo Smerlak}\email{smerlak@cpt.univ-mrs.fr}\affiliation{Max-Planck-Institut f\"ur Gravitationsphysik (Albert-Einstein-Institut)\\Am M\"ulenberg 1, D-14476 Golm, Germany}

\date{\small\today}

\begin{abstract}\noindent
Diffusive transport is characterized by the scaling law $(\textrm{length})^{2}\propto(\textrm{time})$. In this paper we show that this relationship is significantly altered in curved \emph{analogue spacetimes}. This circumstance provides an opportunity to \emph{tailor} diffusion: by a suitable design of the analogue metric, it is possible to create materials where diffusion is either \emph{faster} or \emph{slower} than in normal media, as desired. This prediction can in principle be tested experimentally with optical analogues, curved graphene sheets, etc. -- indeed with any analogue spacetime.

\end{abstract}

\maketitle



\section{Introduction}

Diffusion phenomena are ubiquitous. They underlie thermal and electrical conduction, turbidity, osmosis, breathing, but also sound propagation in rooms \cite{Ollendorff1969,.Picaut1997} and streets \cite{Picaut1999}, opinion spreading in crowded halls \cite{Helbing2010} -- and so forth. In these sundry systems, the characteristic feature of the diffusive behaviour is the \emph{diffusion scaling law}: the spatial spreading of a local impulse grows like the \emph{square root} of time.


Does this scaling still hold in a \emph{curved spacetime}? At first sight, this question may seem rather academic, given that spacetime curvature is strong only in extreme astrophysical (dense stars) and cosmological (early universe) situations. We believe it is not. As is now well-understood, a variety of materials effectively behave as \emph{analogue spacetimes}. This means that, as far as low energy excitations are concerned, the effect of microscopic interactions can be accounted for by a non-trivial Lorentzian metric. Examples of such materials include corrugated graphene sheets \cite{Cortijo2007}, Bose-Einstein condensates \cite{Barcelo2001}, metamaterials \cite{Cai2009}, quantum liquids \cite{Volovik2001}, etc. (See \cite{Barcelo2011} for an updated review of analogue gravity.) To study transport phenomena in these systems, it is important to have a good grasp on the effect of spacetime curvature on diffusion, and notably on the diffusion scaling law. 

One way to address this problem is via Einstein's \emph{stochastic approach} to diffusion, exposed in his 1905 paper on Brownian motion \cite{Einstein1905}. Walking in these footsteps, we derived in \cite{Smerlak2011b} the equation governing the probability density of Brownian motion in the effective geometry
\be\label{geom}
ds^{2}=-N^{2}(t,x)dt^{2}+q_{ab}(t,x)dx^adx^b,
\ee
where $(t,x)$ are coordinates comoving with the medium, $N$ is the so-called \emph{lapse function}, and $q_{ab}$ a $D$-dimensional Riemannian metric describing the intrinsic geometry of the spatial slices. We found that, if $q_{ab}=q_{ab}(x)$ is static, the generalized diffusion equation reads \cite{Smerlak2011b}
\be\label{modifieddiffusion}
\pp_{t}p=\kappa\Delta_{q}(Np).
\ee
Here $p$ is the probability density, $\kappa$ the diffusivity and $\Delta_{q}$ the Laplace-Beltrami operator associated to the spatial metric $q_{ab}$ (see below for definitions). Using this equation, it is only a computational matter to obtain the \emph{mean squared displacement} (MSD) as a function of time, and therefore to obtain the curvature corrections to the diffusion scaling law. It is the purpose of this letter to present our solution to this problem.  

What makes this computation particularly interesting is the fact that the metric coefficients $N$ and $q_{ab}$ are often \emph{tunable} in gravitational analogues. For instance, in optical media the lapse function $N$ is nothing but the inverse refractive index, which recent techniques (metamaterials, non-linear Kerr effect) allow to design very efficiently. This circumstance suggests that diffusive transport can perhaps be \emph{tailored} in analogue spacetimes: either enhanced or slowed down -- at will. By analogy with the current research on metamaterials \cite{Cai2009}, where the propagation of electromagnetic radiation can be tailored with an effective spacetime metric, one can perhaps describe our result as establishing the possibility of ``\emph{metadiffusion}''. 


For clarity, we will investigate this issue by considering separately the possibility of \emph{temporal} and \emph{spatial} tailoring of the MSD, viz. by distinguishing the cases where $N$ and $q$ are only functions of time (resp. space). Once these two cases are understood, it is an easy matter to treat the general situation, where $N$ and $q$ depend on both time and space. 

Among the possible behaviors for the MSD as a function of time, we will find:
\begin{itemize}
\item
a diffusive-to-ballistic crossover in hyperbolic spatial geometries,
\item
an exponential growth with a parabolic lapse profile (``Maxwell's fish eye''),
\item
a finite limit as $t\rightarrow\infty$ in the presence of an analogue event horizon,
\item
any function $f(t)$ such that $f(0)=0$ with a time-dependent lapse function.
\end{itemize}
At the very least, this variety of behaviors demonstrates that diffusion in curved spacetimes is an interesting topic in its own right. We will comment on one possible application in the conclusion. 

\section{Definitions and assumptions}\label{def}

Consider a diffusion process within an in \emph{irrotational} fluid flow. In the instantaneous rest frame (Lagrangian, or comoving coordinates), the effective spacetime metric can be written as
\be\label{geom}
ds^{2}=-N^{2}(t,x)dt^{2}+q_{ab}(x)dx^adx^b
\ee
where $N$ is the \emph{lapse function}, and $q_{ab}$ a static $D$-dimensional Riemannian metric describing the intrinsic geometry of the constant-$t$ spatial slices $\Sigma_{t}$.

Fix a spatial point $x_{0}\in\Sigma_{0}$, and consider the Green function (heat kernel) $K_{t}(x,x_{0})$ of the generalized diffusion equation \eqref{modifieddiffusion}, viz. the solution with initial condition 
\be
\u{t\rightarrow0}{\lim}\ K_{t}(x,x_{0})=\delta(x,x_{0}),
\ee
where $\delta(x,x_{0})$ is the Dirac distribution on the $t=0$ spatial slice $\Sigma_{0}$. Hereafter, we will denote $\langle T,\phi\rangle_{t}$
the pairing between a distribution $T$ and a test function $\phi$ on $\Sigma_{t}$, so that e.g. $\langle\delta,\phi\rangle_{t}=\phi(t,x_{0})$. 

To avoid dealing with drift effects, we will assume furthermore that $q_{ab}$ is spherically symmetric about $x_{0}$, viz.
\be
q_{ab}(\rho)dx^adx^b=a^{2}(\rho)d\rho^{2}+\rho^{2}d\Omega^{2}_{D-1}.
\ee
Here the radial coordinate $\rho$ has the interpretation $\rho=(A/4\pi)^{1/2}$, where $A$ is the area of the $t=\textrm{const.}$, $\rho=\textrm{const.}$ surface, and $d\Omega^{2}_{D-1}$ is the metric on a unit $(D-1)$-sphere centered on $x_{0}$. Without loss of generality, we take $N(0,x_{0})=1$. 

With these assumptions, the squared distance between $x$ and $x_{0}$ at time $t$ is given by
\be
d^{2}(x,x_{0})=\int_{0}^{\rho(x)}d\rho'\,a(\rho'),
\ee
and we can define the MSQ by
\be
\langle d^{2}\rangle_{t}=\langle K_{t},d^{2}\rangle_{t}.
\ee
In a flat spacetime, where $N=a=1$, the MSQ is well-known to be given by $\langle d^{2}\rangle_{t}=2\kappa Dt$: this is the normal diffusion scaling law. In the more general metric considered in this paper, however, such a linear behavior is valid only in the $t\rightarrow0$ asymptotic regime
\be
\langle d^{2}\rangle_{t}\u{t\rightarrow0}{\sim} 2 \kappa  Dt.
\ee
At later times, corrections depending on $N$ and $a$ are to be expected, since the Brownian particle then has had enough time to explore its neighborhood, and felt the effects of its non-trivial geometry. We call \emph{tailoring of diffusion} the possibility of tuning $\langle d^{2}\rangle_{t}$ as a function of $t$ by means of the parameters $N$ and $a$.

\section{Temporal tailoring}\label{temporal}
Let us consider first the possibility of \emph{temporal tailoring}, by which we mean that $N=N(t)$, and $a=1$. In this case, the equation for the Green function simplifies to 
\be\label{temporal1}
\pp_{t}K_{t}=N\kappa\Delta K_{t},
\ee
where $\Delta$ is the standard (time-independent) Laplace operator. Defining
\be\label{defstemporal}
s(t)=\int_{0}^t dt' N(t')
\ee
and performing the change of variables $t\mapsto s(t)$ in equation \eqref{temporal1}, we get
\be
\pp_{s}K_{s}=\kappa\Delta K_{s}.
\ee 
This is nothing but the standard diffusion equation, hence
\be\label{tempMSD}
\langle d^{2}\rangle_{t}=2\kappa Ds(t).
\ee
This result implies that, with a suitably designed $N$, one can tailor the MSD to \emph{any} desired function $f(t)$ such that $f(0)=0$. Indeed, differentiating $\langle d^{2}\rangle_{t}=f(t)$ with respect to $t$ and using \eqref{tempMSD}, we obtain 
\be
N=\f{f'}{2\kappa D}.
\ee
A lapse function satisfying this condition will ensure that the MSD matches the desired function $f(t)$.
\section{Spatial tailoring}\label{spatial}

Even more intriguing is the possibility of \emph{spatial tailoring} with a static, inhomogeneous spacetime metric:

\be
ds^{2}=-N^{2}(\rho)dt^{2}+a^{2}(\rho)d\rho^{2}+\rho^{2}d\Omega^{2}_{D-1}.
\ee
With this spacetime geometry, we cannot in general compute the MSD in closed form. However, we can easily obtain its $t\rightarrow0$ asymptotic expansion, as follows. 

From now on, denote $\mathcal{L}$ the differential operator $\Delta_{q}(N\,\cdot\,)$, so that the generalized diffusion equation \eqref{modifieddiffusion} reads
\be
\pp_{t}K_{t}=\kappa\mathcal{L}K_{t}.
\ee 
This equation can be solved formally as
\be
K_{t}(x,x_{0})=e^{\kappa t\mathcal{L}}\delta(x,x_{0}),
\ee
or, expanding the exponential in powers of its argument,
\be
K_{t}(x,x_{0})=\sum_{n=0}^{\infty}\f{(\kappa t)^{n}}{n!}\mathcal{L}^{n}\delta(x,x_{0}).
\ee
Now, for any test function $\phi$, we have
\be
\langle \mathcal{L}\delta,\phi\rangle_{t}=\langle \delta,\mathcal{L}^{\dagger}\phi\rangle_{t}=(\mathcal{L}^{\dagger}\phi)(t,x_{0}).
\ee
where $\mathcal{L}^{\dagger}=N\Delta_{q}$ is the formal adjoint of $\mathcal{L}$. Using this relationship, we find
\be\label{expansion}
\langle d^{2}\rangle_{t}=\sum_{n=0}^{\infty}\f{(\kappa t)^{n}}{n!}\left((\mathcal{L}^{\dagger})^{n}d^{2}_{t}\right)(x_{0},x_{0}).
\ee
Explicitly, the order-$n$ coefficient $d^{2}_{n}$ of the Taylor expansion $\langle d^{2}\rangle_{t}=\sum_{n=0}^{\infty}d^{2}_{n}t^{n}$ is therefore
\be\label{taylor}
d^{2}_{n}=\f{\kappa^{n}}{n!}\Big(\underbrace{N\Delta_{q}\big(\cdots(N\Delta_{q}}_{n\ \textrm{times}}d^{2})\big)\Big)(x_{0},x_{0}).
\ee

Finally, observe that the asymptotic expansion \eqref{expansion} can be resummed formally as 
\be
\langle d^{2}\rangle_{t}=\big(e^{\kappa t\mathcal{L}^{\dagger}}\big)d^{2}(x_{0},x_{0}).
\ee
This means that the MSQ itself can be computed from a partial differential equation, namely as $u_{t}(x_{0})$, where $u_{t}(x)$ is the solution to the equation
\be\label{backward}
\pp_{t}u_{t}=\kappa\mathcal{L}^{\dagger}u_{t}
\ee
with initial condition $u_{0}(x)=d^{2}(x,x_{0})$. This last equation is referred to as the backward Kolmogorov equation associated to \eqref{modifieddiffusion} in the mathematical literature. It can be useful to compute the MSD numerically.


\section{Two examples of spatial tailoring}\label{examples}

To illustrate these findings, let us consider two cases of spatial tailoring of direct relevance for applications: a quadratic lapse profile, and a spatial geometry with constant curvature.

\subsection{Parabolic lapse profile}

Assume first that the spatial sections are Euclidean ($a=1$), and the lapse function is \emph{parabolic}, 
\be
N(\rho)=1+\epsilon\f{\rho^{2}}{\sigma^{2}}. 
\ee
Here $\epsilon=\pm1$ indexes the convexity/concavity of the profile. Such a profile is used e.g. in graded-index optical fibers. If $\epsilon=1$, it is sometimes referred to as \emph{Maxwell's fish-eye}, and has recently attracted a great deal of attention for its \emph{perfect lensing} properties \cite{Leonhardt2009a}. 

From a computational viewpoint, this example is particularly simple, because both $\Delta\rho^{2}$ and $\Delta N$ are constant, given by $2D$ and $2D \epsilon/\sigma^{2}$ respectively. Thus for $n\geq1$ we have
\be
(\mathcal{L}^{\dagger})^{n}\rho^{2}=\Delta\rho^{2}(\Delta N)^{n-1}=\f{(2D)^{n}\epsilon^{n-1}}{\sigma^{2(n-1)}}.
\ee
It follows that
\be
\langle d^{2}\rangle_{t}= \epsilon\sigma^{2}\sum_{n=1}^{\infty}\f{(\kappa t)^{n}}{n!}\left(\f{2D \epsilon}{\sigma^{2}}\right)^{n}
\ee
i.e.
\be
\langle d^{2}\rangle_{t}=\epsilon \sigma^{2}\left(\exp\Big(\f{2\kappa \epsilon Dt}{\sigma^{2}}\Big)-1\right).
\ee
The interpretation of this result is as follows:

\begin{itemize}
\medskip
\item
If the lapse profile is convex ($\epsilon=1$), diffusion \emph{speeds up} in time, with an exponential growth of the MSD with time scale $t^*=\sigma^{2}/2\kappa D$. With respect to normal diffusive scaling, such a profile is \emph{repulsive}.
\item
 If the lapse profile is concave ($\epsilon=-1$), diffusion \emph{slows down} in time, until it eventually stops at $t\simeq t^{*}=\sigma^{2}/2\kappa D$, when the MSD reaches the finite limit $\sigma^{2}$. With respect to normal diffusive scaling, such a profile is \emph{attractive}.
\end{itemize}
From a relativist's perspective, the stopping of diffusion on the surface $\rho=\sigma$ does not come as a surprise: it is an instance of the ``freezing'' phenomenon typical of event horizons. We conjecture that the same behavior would occur in the vicinity of any \emph{infinite redshift surface}, where the lapse function $N$ vanishes.

\subsection{Pure spatial curvature}

Another interesting special case is when $N=1$, viz. when the only non-trivial curvature components are spatial. In this case, already considered in the biophysical context \cite{Yoshigaki2007,Castro-Villarreal2010}, the order-$n$ Taylor coefficient of the MSD as a function of time \eqref{taylor} reduces to 
\be
d_{n}^{2}=\f{k^{n}}{n!}\Delta_{q}^{n}d^{2}(x_{0},x_{0}).
\ee
The quantities $\Delta_{q}^{n}d^{2}(x_{0},x_{0})$ are well-known geometric invariants, tabulated e.g. in \cite{DeWitt2003}.\footnote{See also for \cite{Ottewill2011} fast numerical algorithms.} The first two read
\begin{eqnarray}
\Delta_{q}d^{2}(x_{0},x_{0})&=&2D,\\
\Delta_{q}^{2}d^{2}(x_{0},x_{0})&=&-\f{4}{3}R_{q}(x_{0}),
\end{eqnarray}
where $R_{q}(x_{0})$ is the spatial Ricci curvature. Hence, to next to leading order, we find that a negative spatial curvature will \emph{enhance} the MSD, while a positive spatial curvature will \emph{diminish} it. The characteristic time of the diffusive-to-ballistic crossover in the hyperbolic case is $t^{*}=D/\kappa R_{q}(x_{0})$. 

%

\section{Comments}

Several comments may be useful to clarify the nature of the results discussed in the preceding sections. 

First, let us stress that the main equation of this paper, namely eq. \eqref{modifieddiffusion}, is not merely the standard diffusion equation written in a non-standard coordinate system. It is a different and more general equation, which reduces to the standard diffusion only when the spacetime curvature and background accceleration both vanish. In the presence of spacetime curvature, as for instance in solids with disclinations or in gradient-index optical media, it leads to genuinely different physical predictions. Furthermore, the coordinate system used to write \eqref{geom} and \eqref{modifieddiffusion} is not arbitrary: it is \emph{the} comoving frame in which the background medium is locally at rest.

Second, the generalized diffusion equation \eqref{modifieddiffusion} is actually well-known in general relativity: it is the \emph{Eckart heat equation} \cite{Eckart1940}. (See \cite{Smerlak2011b} for more details.) Here, thanks to the connection with Brownian motion established in \cite{Smerlak2011b}, it is applied to \emph{any} diffusion phenomenon in the presence of an analogue gravitational field, and not just to heat dynamics in dissipative relativistic fluids.

Third, it is worth emphasizing that the analogue gravity picture of condensed-matter systems underlying this work is only an effective description, which hides a great deal of microscopic physics. This approximation is similar to the one consisting in describing an optical medium by a single scalar field, the refractive index. It is valid only when the relevant excitations, as the photons in the optical case, have low energy and large wavelength and therefore do not couple to microscopic inhomogeneities. Thus, we have assumed in particular that transport can be described as a local diffusion, involving the Laplace-Beltrami operator in space $\Delta_{q}$ (rather for instance than as a diffusion in phase space, as in the Kramers kinetic equation). As usual, this implies that the relevant mean free path is small compared to other length scales in the problem. 

Fourth, the Green function $K_{t}$ of the generalized diffusion equation \eqref{modifieddiffusion}, unlike its flat-spacetime equivalent, is in general not Gaussian. This implies in particular that $K_{t}$ is \emph{not} fully characterized by the associated MSD. For this reason, it will be necessary to push further the analysis initiated in this paper in order to get a comprehensive picture of ``diffusion tailoring''. One way to gather further information about the effect of analogue gravity on diffusion is to compute the time-evolution of the entropy of $K_{t}$ as a function of the effective geometry; this task will be undertaken in a future publication.

\section{Conclusion and outlook}\label{conclusion}

In this paper we have computed the MSD for Brownian motion in a curved, spherically symmetric, analogue spacetime. Depending on the convexity of the lapse function and the sign of spatial curvature, we have found that the MSD can either grow \emph{faster} or \emph{slower} than in flat spacetime. In the extreme case where the analogue spacetime geometry presents an infinite redshift surface ($N=0$), diffusion actually \emph{stops}. In principle, any one of the many condensed-matter systems behaving as analogue spacetimes can be used to test these predictions in the lab. 


Diffusion is such a common phenomenon in physics, chemistry, etc., that it is impossible to list all the possible applications of ``metadiffusion''. Let us simply mention one, 
related to Anderson localization.\footnote{I thank Daniele Faccio for suggesting this example to me.} According to the so-called scaling theory, the critical point of the localization transition in a finite sample of size $L$ can be estimated by the criterion $g(L)\simeq 1$, where $g(L)$ is the ratio of the uncertainty of the energy levels due to the boundary conditions to the mean energy spacing. The former is inversely proportional to the \emph{Thouless time} $T_{L}$: the time needed for a diffusive carrier to reach the boundaries of the sample. In normal situations, this time is given by $T_{L}=\kappa/L^{2}$. In a sample with curved effective geometry, however, we have found that this relationship can be modified, and actually tailored. Can this be used to control the localization threshold of analogue spacetimes? This question will be addressed in a future work.


\smallskip

\emph{Acknowledgements.} I am grateful to Daniele Faccio and Valentin Bonzom for discussions and feedback on the ideas presented in this paper.  

\medskip 
\emph{Note added in proof. The concept of ``diffusion tailoring'' was also discussed in the context of heat dynamics in \cite{Guenneau2012}, which appeared after the first submission of this article to Phys. Rev. E.}

\bibliographystyle{utcaps}\bibliography{library}


\end{document}